\documentclass[preprint,12pt]{elsarticle}

\usepackage{latexsym}
\usepackage{amsfonts}
\usepackage{color}
\usepackage{graphicx}
\usepackage{mathptmx}      
\usepackage{bm}
\usepackage{enumerate}
\usepackage{subfigure}
\usepackage{listings}
\usepackage{grffile}
\usepackage{multirow}
\definecolor{DarkGreen}{rgb}{0.00,0.39,0.00}
\usepackage{rotating,tabularx}

\newcommand\IBMSTATE[5]{|#1_4 #1_3 #2_2 #3_1 #4_0\rangle}
\newcommand\CSTATE[3]{|#1_2 #2_1 #3_0\rangle}
\newcommand\STATE[4]{|#1_3 #2_2 #3_1 #4_0\rangle}
\newcommand\ASTATE[4]{|#1_4 #2_3 #3_2 #4_1\rangle}
\newcommand\state[4]{|#1_{#2} #3_{#4}\rangle}

\journal{Computer Physics Communications}

\begin{document}

\begin{frontmatter}



\title{Benchmarking gate-based quantum computers}


\author[FZJ,RWTH]{Kristel Michielsen\footnote{Corresponding author: k.michielsen@fz-juelich.de}}
\author[FZJ]{Madita Nocon}
\author[FZJ]{Dennis Willsch}
\author[FZJ]{Fengping Jin}
\author[FZJ]{Thomas Lippert}
\author[RUG]{Hans De Raedt}

\address[FZJ]{Institute for Advanced Simulation, J\"ulich Supercomputing Centre, Forschungzentrum J\"ulich, D-52425 J\"ulich, Germany}
\address[RWTH]{RWTH Aachen University, D-52056 Aachen, Germany}
\address[RUG]{Zernike Institute for Advanced Materials, University of Groningen, Nijenborgh 4, NL-9747 AG Groningen, The Netherlands}

\begin{abstract}
With the advent of public access to small gate-based quantum processors,
it becomes necessary to develop a benchmarking
methodology such that independent researchers
can validate the operation of these processors.
We explore the usefulness of a number of simple quantum circuits as
benchmarks for gate-based quantum computing devices and show
that circuits performing identity operations are very simple,
scalable and sensitive to gate errors
and are therefore very well suited for this task.
We illustrate the procedure by presenting benchmark results
for the IBM Quantum Experience, a cloud-based platform
for gate-based quantum computing.
\end{abstract}



\end{frontmatter}

\section{Introduction}\label{section0}

As small gate-based quantum computer hardware is being made available to the public~\cite{BRISTOL,IBMQE},
it is now possible for independent parties to validate and benchmark the operation of these devices.
Therefore, it seems natural to introduce a suite of quantum algorithms (i.e. sequences of gate operations~\cite{NIEL10})
which should be used to validate quantum processors.
The aim of this paper is to explore the potential of several different, simple sequences of gate operations
that can be used for this task, building on earlier work that was specifically targeting NMR quantum processors~\cite{RAED02}.

A gate-based quantum computer is a device that takes input data and transforms this input data according to a unitary operation,
specified as a sequence of gate operations and measurements (i.e. the algorithm)
and conveniently represented by a quantum circuit~\cite{NIEL10}.
The algorithm itself does not depend on the input data and returns the result of the transformation in the form of output data.
If the transformation involves random processes, the output data have to be interpreted according to the probabilistic
model (i.e. quantum theory in the case of a quantum computer) of these random processes.
Evidently, this cursory description of a gate-based quantum computer refers to the highly abstract {\bf mathematical model}
of the device only.

The central question is to what extent the hardware implementation of a quantum processor operates according to the
mathematical model and can therefore deliver the exponential speed-up that this mathematical model promises~\cite{NIEL10}.
Although one can think of many physical processes that cause the hardware implementation to function
in a way that differs from the one imagined on the basis of its circuit model~\cite{NIEL10}, from
a user perspective (but not from the perspective of the manufacturer),
it is immaterial whether a malfunctioning of the device can be attributed to a particular physical process or not.
The only thing that matters is whether the device performs the desired computation properly.

For a device that performs the mapping ``algorithm(input data) $\rightarrow$ output data''
to qualify as a computer, the following two requirements seem essential:

\begin{itemize}
\item
For each instance of the input data and with the algorithm fixed,
the relation algorithm(input data) $\rightarrow$ output data should yield (within statistical fluctuations)
the correct output data.
In the case of a gate-based quantum computer, the correct output data can be obtained
by running the algorithm on the mathematically exact, pen-and-paper-model of the quantum computer.
For this purpose, one can use a massively parallel quantum computer simulator~\cite{RAED07x,ITO17}
running on PC's or supercomputers such as JUQUEEN~\cite{JUQUEEN} (an IBM Blue Gene/Q) or the K computer,
allowing the simulation of up to 45 qubits (on the K computer).
If the number of qubits does not exceed 5 it is more convenient
to execute the algorithm on the simulator included in the IBM Quantum Experience (IBM-QE)~\cite{IBMQE}.
\item
For the same input data and with the algorithm fixed,
the output data should be stationary in time.
Disregarding statistical fluctuations, this means that the output data should not change
with the time or day when the procedure is carried out.
\end{itemize}

In this paper, we strictly take the viewpoint of a potential user of a quantum processor,
that is we explore the use of simple but decisive gate sequences to check if the hardware implementation of a quantum processor
complies with the two aforementioned requirements for being a useful computing device.
We illustrate the procedure by running these tests on the IBM-QE~\cite{IBMQE}
and demonstrate that this device does not qualify accordingly.
We assume that the reader has some elementary notion of what quantum computation is about but again,
from the viewpoint of a user 
such knowledge is not required to properly interpret the results of the tests that we present.

The paper is organized as follows.
In section~\ref{section1}, we give a brief overview of the IBM-QE hardware, confining
ourselves to those aspects that are relevant for the user who wants to run applications on the processor.
We also discuss the procedure of collecting and analyzing the experimental data.
Section~\ref{section2} presents results for a very simple but instructive application, the preparation
and subsequent measurement of the singlet state.
Section~\ref{section3} explores the potential of using a two-register adder as a check on the hardware.
The corresponding quantum circuit~\cite{DRAP00} involves the quantum Fourier transform~\cite{NIEL10} and
performs addition modulo 4. The adder circuit has the appealing feature that it is trivial
to check whether or not the hardware does the addition correctly.
Moreover, as it requires significantly more gate operations than the circuit employed
in section~\ref{section2}, it may be expected to be more prone to the accumulation of errors.
Section~\ref{section4} introduces a very simple, flexible and scalable class
of quantum circuits that prove to be well-suited for validating quantum processors.
The key is to perform identity operations or, in other words, no operation at all.
Within the mathematical model of the quantum processor,
each gate operation corresponds to a unitary transformation on the qubits and hence
it is almost trivial to construct sequences of identity operations.
Also in this case, it is easy to decide whether the processor functions properly or not.
In section~\ref{section5}, we scrutinize the usefulness of two different error correction
schemes and show that in practice, meaning on the IBM-QE hardware, these schemes do not live up to the expectations, namely instead
of reducing they enhance the chance for an incorrect result.
Section~\ref{section6} contains a discussion of the conclusions that we draw on the basis
of the experimental data.

\section{The IBM quantum experience}\label{section1}

Since May 2016 IBM has been providing public access to a 5-qubit quantum processor~\cite{IBMQE}.
The first version of this processor allowed for single-qubit operations on all qubits
and CNOT operations between 4 qubits (numbered 0,1,3,4) and the qubit number 2,
i.e. between the qubit pairs (0,2), (1,2), (3,2), and (4,2) where
the first element of the pairs denotes the control qubit.
The present version allows for additional CNOT operations, namely between the pairs (0,1) and (3,4).
The device is accessible through a web interface which provides the necessary tools
to execute quantum programs on the device, as well as on a simulator that
performs the operations according to the mathematical model of the idealized device.

An experiment on the IBM-QE consists of
(i) specifying the quantum circuit, either through a graphical
interface or a text-based editor, (ii) running the circuit on the simulator to check
if the circuit has been specified correctly, and (iii) executing the circuit on the hardware processor
for a number $N$ of so-called ``shots''.
With each shot, the processor is first initialized and is then instructed (by a controller not accessible to the user)
to execute the quantum circuit.

Barriers prevent software optimization of successive gates in a circuit~\cite{IBMQE}.
As this optimization was turned off at the time we did our experiments,
there was no need to include barriers in our circuits.
However, to make sure that the IBM-QE software does not optimize the circuit,
we recommend including barriers in future experiments.

The final state of the device is read out by a process called ``measurement'' which,
for each shot, returns either the values 0 or 1 for each of the measured qubits.
Thus, each shot yields one string of at most 5 bits which may or may not be different each time the circuit is executed.
After $N$ shots, the system returns the counts of the number of times that each of the different bit strings was generated.
These counts, not the individual bit strings, constitute the result of executing the quantum algorithm.

\subsection{Device characteristics}\label{section1b}

According to the IBM-QE documentation and private communication with the IBM-QE team,
execution of an X, Hadamard, and CNOT gate takes $130\;$ns, $130\;$ns, and $650\;$ns, respectively.
The coherence time of a single qubit is of the order of $100\;\mu\mathrm{s}$.
The gate errors, estimated from randomized benchmarking~\cite{CHOW09}, are in the range $10^{-2}$ -- $10^{-3}$.
These are parameters of the device that was in use between January 11, 2017 and February 6, 2017.
These numbers vary somewhat from one device calibration to another (typically twice a day).

\subsection{Data analysis}\label{section1a}
Each run of an algorithm on the IBM-QE yields a definite pattern of (five) output bits (0 or 1).
This pattern, a basis state in quantum theory parlance,
is interpreted as being the result of measuring the quantum state of the machine.
We denote a basis state of the IBM-QE by $\IBMSTATE{Q}{Q}{Q}{Q}{Q}$ where the Q's are either 0 or 1,
and we use a similar notation for less than 5 qubits.
The subscripts correspond to the labels of the qubits of the IBM-QE and are necessary
because some algorithms might permute logical and physical qubits, as in the case of the adder
and the error correction algorithm discussed below.

In this paper, we present results of repeating the procedure of executing the algorithm and measuring
the state of the device $N=8192$ times, the maximum number of shots currently allowed.
The counts of the different configurations of 0's and 1's
divided by the number of shots give us the relative frequencies (rational numbers between 0 and 1, with
their sum being equal to one) with which the basis states are observed.

\begin{center}
\framebox{
\parbox[t]{0.8\hsize}{%
In this paper, we take as the ``correct'' result of the computation either
the collection of states with the largest relative frequencies
or
the state that appears with the largest relative frequency if a unique answer is expected.
}}
\end{center}

If quantum theory is assumed to describe the operation of the device,
each measurement constitutes a statistically independent trial~\cite{BALL03}.
Within statistical fluctuations, the measured frequencies to observe the system in one of the basis states
should correspond to the probabilities predicted by quantum theory~\cite{BALL03}.
Let us denote by $x_n=1(0)$ the fact that a particular basis state is (not) observed in trial $n=1,\ldots,N$.
The relative frequency is then $f=N^{-1}\sum_{n=1}^N x_n$.
Assuming that the $x_n$ are identically distributed random variables,
the standard error (SE) on the estimated value of $f$ is bounded by $1/\sqrt{N}$.
For the case at hand, we conclude that {\bf if the measurements constitute identically distributed random trials},
the standard error on the data that we present in this paper (with $N=8192$) does not exceed $\mathrm{SE}=0.012$.
Frequencies that are within five standard errors ($5\;\mathrm{SE}=0.06$) are considered to be the same.

It is to be expected that the results obtained by executing a quantum circuit not only
suffer from statistical errors but, as shown below, also from other errors that are much harder to characterize properly.
Let us assume that for each gate operation,
there is a probability $0 < p_\mathrm{C} <1$ that the result of the operation,
if measured, is correct and that the probability for a sequence of $m$ identical gates
to return the correct answer is $ p_\mathrm{C}^m$.
For instance, taking $p_\mathrm{C}=0.95$, this error model would predict
a sequence of $m=20$ gates to produce the correct result with a probability of about 0.36, which is larger
than the probability for sampling (uniformly) at random if the number of qubits is larger than one.
Our experiments strongly suggest that the variations in the experimental results
presented below are much larger and cannot be explained by probabilistic error models based on the single-gate errors,
often estimated from randomized benchmarking~\cite{EMER05}, supporting the viewpoint that the estimates
obtained by the latter have some deficiencies~\cite{PROC17}.

\section{Entanglement}\label{section2}

\begin{figure}[t]
\begin{center}
\includegraphics[width=0.48\hsize]{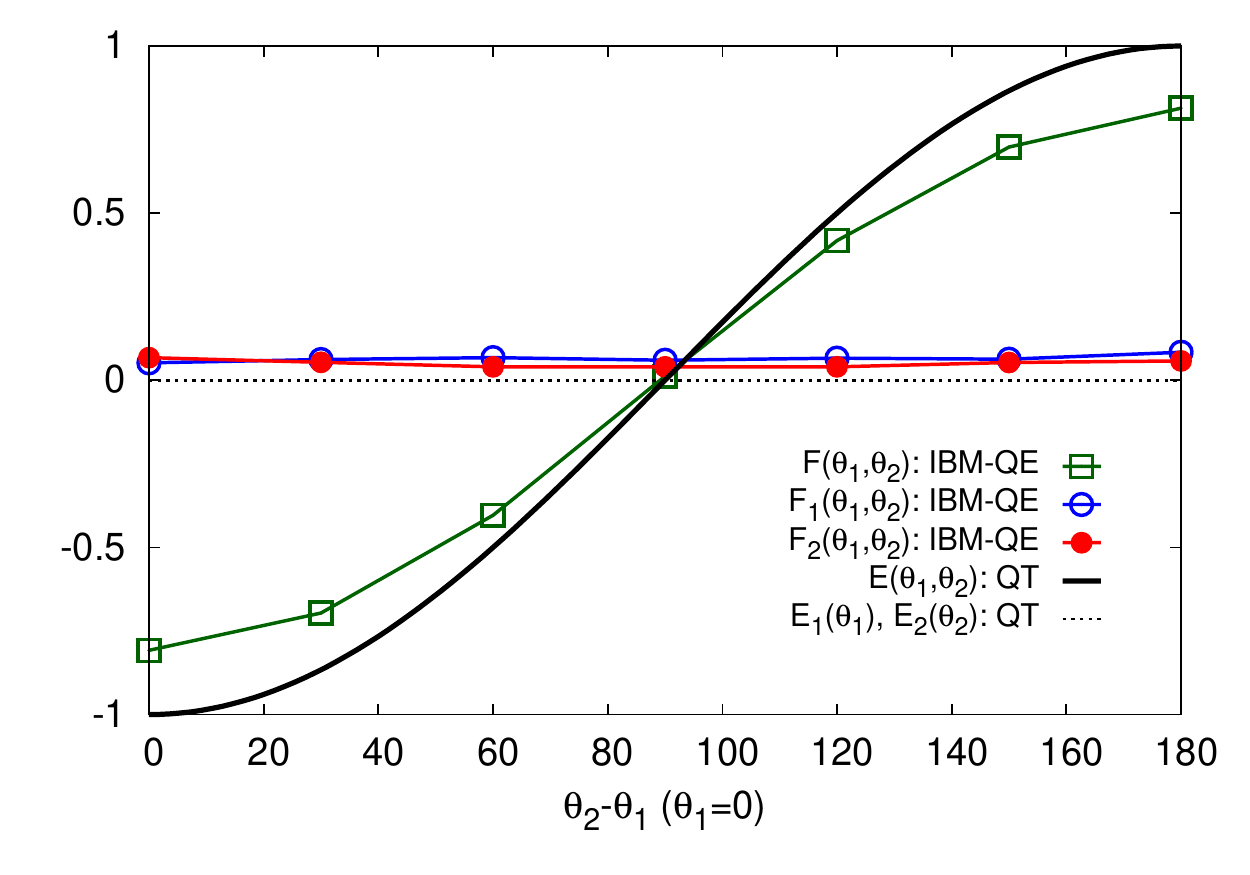}
\includegraphics[width=0.48\hsize]{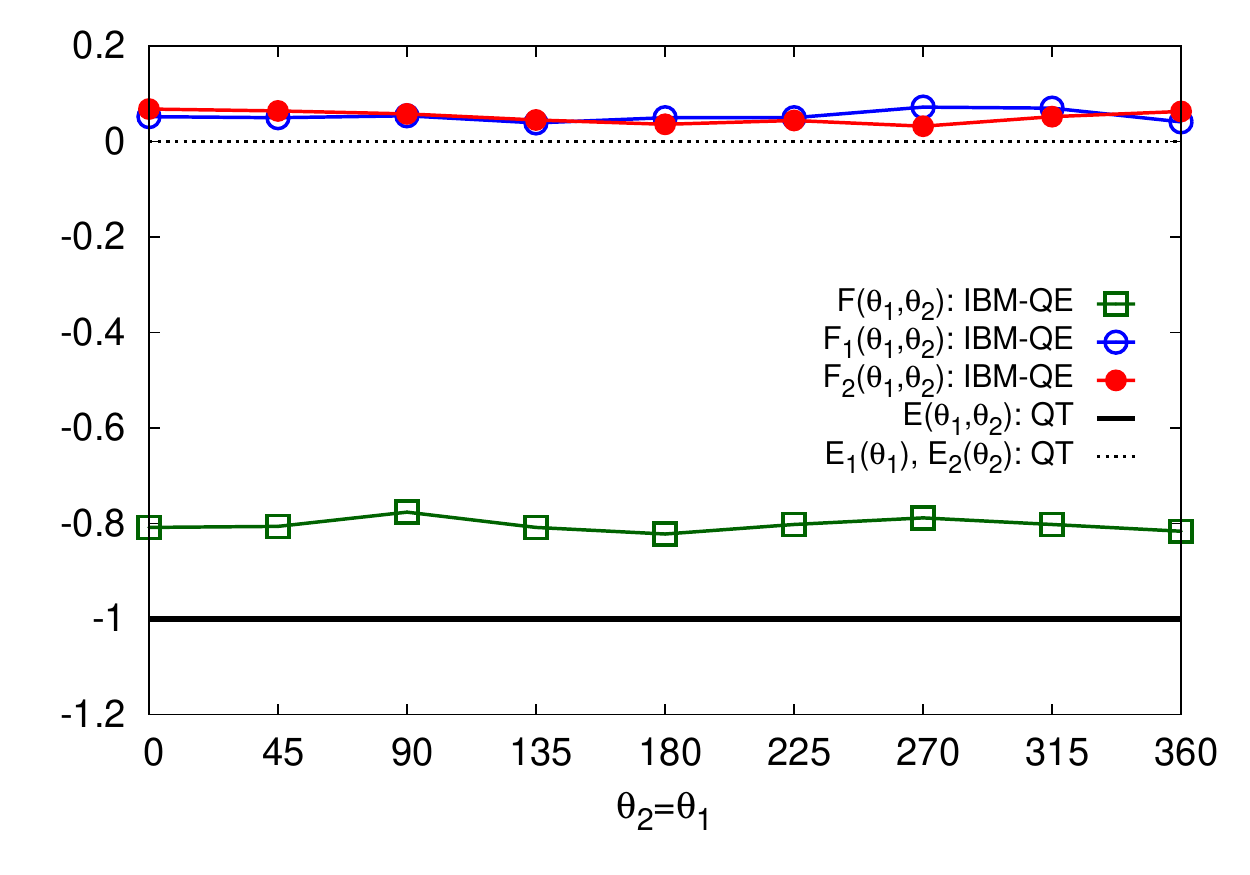}
\caption{(color online) %
Data for a quantum-gate circuit which, theoretically, generates the singlet state.
The quantum-gate circuit used is given in \ref{appA}.
The experiments on the IBM-QE have been carried out on February 16, 2017 with
$\mathbf{a}=(0,-\sin\theta_1,\cos\theta_1)$ and $\mathbf{b}=(0,-\sin\theta_2,\cos\theta_2)$.
Left: $\theta_1=0$ fixed and $\theta_2$ variable, in which case quantum theory predicts $E_1(\theta_1)=E_2(\theta_2)=0$,
$E(\theta_1,\theta_2)=-\cos(\theta_1-\theta_2)$;
right: $\theta_1=\theta_2$ variable, in which case quantum theory predicts $E_1(\theta_1)=E_2(\theta_2)=0$, $E(\theta_1,\theta_2)=-1$.
Lines connecting the data points are guides to the eye.
}
\label{fig4}
\end{center}
\end{figure}

A conceptually simple experiment to test whether a two-qubit system is capable of
exhibiting quantum behavior is to
repeatedly prepare the device such that the readout of its internal state yields
a frequency distribution of events that agrees with the probability distribution
of two spin-1/2 particles in the singlet state, a maximally entangled state~\cite{BALL03}.
Note that the observation that the frequency distribution of many events agrees with
the probability distribution of the singlet state is only a post-factum
characterization of the repeated preparation and measurement process,
not a demonstration that at the end of the preparation stage, the device actually {\sl is} in the singlet state.
The latter describes the statistics, not the internal state at any particular instance~\cite{BALL03}.


The singlet state is defined by
\begin{eqnarray}
|\Psi\rangle&=&\frac{1}{\sqrt{2}}\left( |01\rangle - |10\rangle\right)
.
\label{singletstate0}
\end{eqnarray}%
The averages of (combinations of) Pauli spin matrices $\bm{\sigma}=(\sigma^x,\sigma^y,\sigma^z)$ are
\begin{eqnarray}
\langle\Psi|\bm{\sigma}_1\cdot \mathbf{a} |\Psi\rangle&=&\langle\Psi|\bm{\sigma}_2\cdot \mathbf{b} |\Psi\rangle=0
,
\label{singletstate1a}
\\
E(\mathbf{a},\mathbf{b})&=&
\langle\Psi|\bm{\sigma}_1\cdot \mathbf{a}\;\bm{\sigma}_2\cdot \mathbf{b} |\Psi\rangle=-\mathbf{a}\cdot \mathbf{b}=-\cos\theta
,
\label{singletstate1b}
\end{eqnarray}%
where $\mathbf{a}$ and $\mathbf{b}$ are three-dimensional unit vectors and $\theta$ is the angle between these two vectors.

It is not difficult to see that in general, any function $P(S_1,S_2|\mathbf{a} \mathbf{b})$
of the two-valued variables $S_1,S_2=\pm1$ can be written as
\begin{eqnarray}
P(S_1,S_2|\mathbf{a} \mathbf{b})&=&\frac{P_0 + P_1 S_1 + P_2 S_2 + P_3 S_1 S_2}{4}
,
\label{Psinglet1z}
\end{eqnarray}%
simply because
\begin{eqnarray}
\sum_{S_1=\pm1}\sum_{S_2=\pm1} P(S_1,S_2|\mathbf{a} \mathbf{b}) &=& P_0
\nonumber\\
\sum_{S_1=\pm1}\sum_{S_2=\pm1} S_1 P(S_1,S_2|\mathbf{a} \mathbf{b}) &=& P_1
\nonumber\\
\sum_{S_1=\pm1}\sum_{S_2=\pm1} S_2 P(S_1,S_2|\mathbf{a} \mathbf{b})  &=&  P_2
\nonumber\\
\sum_{S_1=\pm1}\sum_{S_2=\pm1} S_1 S_2 P(S_1,S_2|\mathbf{a} \mathbf{b})  &=& P_3
.
\label{Psinglet12}
\end{eqnarray}%
If $P(S_1,S_2|\mathbf{a} \mathbf{b})$ is to represent the probability that a measurement of the spins
$(\bm\sigma_1\cdot\mathbf{a}, \bm\sigma_2\cdot\mathbf{b})$
yields the values $(S_1,S_2)$, we must have $P_0=1$ and therefore
\begin{eqnarray}
P(S_1,S_2|\mathbf{a} \mathbf{b})&=&\frac{1 + P_1  S_1 + P_2 S_2 + P_3 S_1 S_2}{4}
,
\label{Psinglet1y}
\end{eqnarray}%
with certain restrictions on $(P_1,P_2,P_3)$ because we must also have $0\le P(S_1,S_2|\mathbf{a} \mathbf{b}) \le 1$
for $P(S_1,S_2|\mathbf{a} \mathbf{b})$ to qualify as a probability.

We find the probability for measuring the spins $(S_1,S_2)$ in the singlet state by
combining Eqs.~(\ref{singletstate1a}), (\ref{singletstate1b}) and (\ref{Psinglet1y}), meaning that we set
$P_1=\langle\Psi|\bm{\sigma}_1\cdot \mathbf{a} |\Psi\rangle=0$,
$P_2=\langle\Psi|\bm{\sigma}_2\cdot \mathbf{b} |\Psi\rangle=0$,
$P_3=\langle\Psi|\bm{\sigma}_1\cdot \mathbf{a}\;\bm{\sigma}_2\cdot \mathbf{b} |\Psi\rangle=-\mathbf{a}\cdot \mathbf{b}$
and obtain
\begin{eqnarray}
P_{\mathrm{singlet}}(S_1,S_2|\mathbf{a} \mathbf{b})&=&\frac{1- S_1 S_2\mathbf{a}\;\cdot \mathbf{b}}{4}
.
\label{Psinglet1}
\end{eqnarray}%
The mapping from the spin-1/2 eigenvalues $S_i=+1,-1$
to the qubit values $q_i=0,1$ is given by $q_i=(1-S_i)/2$ (or $S_i=1-2q_i$) for $i=1,2$~\cite{NIEL10}.
Therefore, according to quantum theory, the probability that a measurement of the qubits $(Q_1,Q_2)$
yields the values $(q_1,q_2)$  ($q_1,q_2=0,1$), is given by
\begin{eqnarray}
P(q_1,q_2|\mathbf{a} \mathbf{b})&=&\frac{1-\mathbf{a}\cdot \mathbf{b}
+2\mathbf{a}\cdot \mathbf{b}\;(q_1+q_2)
-4\mathbf{a}\cdot \mathbf{b}\;q_1q_2}{4}
\nonumber \\
&=&\frac{1-(-1)^{(q_1+q_2)}\mathbf{a}\cdot \mathbf{b}}{4}
.
\label{Psinglet}
\end{eqnarray}%


The quantum circuit generating the singlet state is very simple:
assuming that the initial state of the two qubits is $|0_20_1\rangle$,
perform an X operation on qubits 1 and 2 to change the state to $|1_21_1\rangle$,
apply a Hadamard gate on qubit 1, and execute a CNOT operation with qubit 1 (2) as  control (target) qubit.
In our experiments, we have chosen $\mathbf{a}=(0,-\sin\theta_1,\cos\theta_1)$ and $\mathbf{b}=(0,-\sin\theta_2,\cos\theta_2)$
such that $\mathbf{a}\cdot \mathbf{b}=\cos(\theta_1-\theta_2)$.
The circuit that, in the ideal case,  generates $(q_1,q_2)$ according to Eq.~(\ref{Psinglet})
and implements the measurement in a rotated basis specified
by $\mathbf{a}$ and $\mathbf{b}$,
then consists of two X-gates, five Hadamard gates, one $U_1(\theta_1)$ and one $U_1(\theta_2)$ gate, and a CNOT gate.
The sequence of gates that implements this circuit is given in \ref{appA}.

Executing the sequence of gates on the IBM-QE yields, after $N=8192$ shots,
the relative frequencies $f(q_1,q_2)$ ($f(0,0)+f(0,1)+f(1,0)+f(1,1)=1$) with which the pair $(q_1,q_2)$ is generated.
The averages and correlation of the Pauli spin matrices projected onto the directions of measurement are given by
$F_1(\theta_1,\theta_2)=f(0,0)+f(0,1)-f(1,0)-f(1,1)$,
$F_2(\theta_1,\theta_2)=f(0,0)-f(0,1)+f(1,0)-f(1,1)$, and
$F(\theta_1,\theta_2)=f(0,0)-f(0,1)-f(1,0)+f(1,1)$, respectively.

For a two-qubit system, the averages of the Pauli-spin matrices are
$E_1(\theta_1)=\langle\Psi|\bm{\sigma}_1\cdot \mathbf{a}|\Psi\rangle$,
$E_2(\theta_2)=\langle\Psi|\bm{\sigma}_2\cdot \mathbf{b}|\Psi\rangle$, and
$E(\theta_1,\theta_2)=\langle\Psi|\bm{\sigma}_1\cdot \mathbf{a}\;\bm{\sigma}_2\cdot \mathbf{b} |\Psi\rangle=-\mathbf{a}\cdot \mathbf{b}$.
Therefore, if the operation of the IBM-QE is described by the quantum theory of a system of qubits, we expect to find that
$F_1(\theta_1,\theta_2)\approx E_1(\theta_1)$,
$F_2(\theta_1,\theta_2)\approx E_2(\theta_2)$, and
$F(\theta_1,\theta_2)\approx E(\theta_1,\theta_2)$.

In Fig.~\ref{fig4}, we present experimental data as obtained by executing the sequence of gates on the IBM-QE.
Qualitatively, the data presented in Fig.~\ref{fig4} show the features that
are expected from the quantum theoretical description in terms of the singlet state
but quantitatively, there are significant deviations.
For $\mathbf{a}=\mathbf{b}=(0,-\sin\theta_1,\cos\theta_1)$ quantum theory predicts that
$\langle\Psi|\bm{\sigma}_1\cdot \mathbf{a}\;\bm{\sigma}_2\cdot \mathbf{b} |\Psi\rangle$
is constant in the range $[0.945,1]$ whereas the experiment (see Fig.~\ref{fig4} (right))
gives $F(\theta_1,\theta_1)\approx0.80$, far outside the expected interval.
The single-qubit averages $F_1(\theta_1,\theta_1)$ and $F_2(\theta_1,\theta_1)$ are zero within a 6 SE margin.
Assuming that the data is described by quantum theory in terms of a pure state,
a nonlinear fit to all the data presented in Fig.~\ref{fig4} yields
\begin{eqnarray}
|\Phi_{\mathrm{data}}\rangle&=& 0.95|\Psi\rangle+ 0.31\frac{|00\rangle - |11\rangle}{\sqrt{2}}+\ldots
.
\label{Psingleta}
\end{eqnarray}%

It is noteworthy that the mentioned artifacts are not only found in the data produced by the IBM-QE experiments
but are also present in data collected in Einstein-Podolsky-Rosen-Bohm experiments with photons~\cite{RAED13a}.
From a more general perspective, it is instructive to compare the ``accuracy'' of the results produced by quantum physics experiments
such as those (but not only those) performed on the IBM-QE with those on e.g. atomic systems.
The accuracy by which quantum theory predicts, say, the ratios of the wavelengths of the Balmer
absorption/emission lines of hydrogen is about 4 digits.
Some of the wavelengths of these lines have been measured with roughly this precision
in the beginning of the previous century with, for present-day standards, pre-historic equipment.
Taking the experiments on entanglement, which involve only a few gate operations, as an example,
experiments (not only the IBM-QE but also experiments with photons, neutrons, ions, ...)
reproduce the quantum-theoretical prediction
for the correlation $\langle\Psi|\bm{\sigma}_1\cdot \mathbf{a}\;\bm{\sigma}_2\cdot \mathbf{a} |\Psi\rangle=-1$
with an accuracy of not more than 2 digits.
Apparently, it seems rather challenging for humans to engineer devices that
operate according to the laws of quantum theory with a precision akin to that of atomic systems found in nature.


\section{Two-qubit + two-qubit adder}\label{section3}

A rather simple but nontrivial algorithm to test the correctness of quantum computer simulation software and hence also devices
is to perform integer addition, which has the appealing feature that it is trivial to check the correctness of
the results generated by the software or device~\cite{RAED07x}.
The algorithm that we use here makes use of a quantum Fourier transform~\cite{DRAP00}.
Due to the limitations of the IBM-QE hardware, the integers that can be added are rather small ($\le 3$).
Nevertheless, running the algorithm on the IBM-QE reveals some interesting behavior, also see the supplementary material of Ref.~\cite{DEVI16}.
The implementation of the adder circuit is different from the one reported in Ref.~\cite{DEVI16}.
It has been validated by running the quantum algorithm on the simulator and comparing the results with those of integer arithmetic modulo 4.

In Table~\ref{tab2a}, we collect a number of cases for which the IBM-QE results are sometimes correct and sometimes wrong.
We have not been able to detect any systematics in this behavior.
QE results that are correct on a particular day may turn out wrong on another date, or vice versa.
See also Ref.~\cite{DEVI16}.

In Table~\ref{tab2b}, we present some data for the case that the inputs are superpositions of the states that represent the integer numbers.
In all cases shown, execution of the algorithm on the IBM-QE returns the expected answer.


\section{Identity operations}\label{section4}

Sequences of several CNOT operations provide simple but decisive test cases~\cite{RAED02}.
Theoretically, each pair of CNOT gates acts as an identity operation,
hence, if the number of CNOT operations is even, we expect to see that the output state is the same as the input state.
In Table~\ref{tab1}, we present some representative results for sequences of 8 and 12 CNOT gates.
Some of these sequences are preceeded/followed by some X and H gates to change the input/output to/of the sequence of CNOTs.
The duration of 12 successive CNOT gates  ($\approx 8\;\mu\mathrm{s}$) is well within the coherence time of the qubits ($100\;\mu\mathrm{s}$).

From Table~\ref{tab1} it is clear that, except for the cases shown in the last two rows, the states that occur with
the largest relative frequency are quite robust: they do not change if we repeat the experiment.
Moreover, all relative frequencies are in the same ball park and fairly large.
However, comparing the data of the second and third row, we must conclude
that there is no guarantee that the device is operating properly.
Obviously, the operation of the device suffers from  errors which are hard to gauge.
We have found no systematic procedure to
determine the conditions under which the device produces blatantly wrong results.



\section{Error correction}\label{section5}

The idea of quantum error correction is to introduce redundancy by using $m$ physical qubits to encode $k<m$ logical qubits such that
the information is effectively protected against decoherence~\cite{shor1995errorcorrection}. Quantum error-correcting
codes are commonly denoted by $[[m,k,d]]$, where the distance $d$ includes information about the number of errors the
code can correct~\cite{NIEL10}. Since the IBM-QE supports five qubits, small codes using $m\le 5$ physical qubits can
be implemented and tested (see also Refs.~\cite{DEVI16} and \cite{takita2016demonstration}).

We study two different five-qubit codes. The first is a particular distance-two surface code already analyzed by Devitt
using a previous version of the IBM-QE~\cite{DEVI16}. The second is an instance of the perfect $[[5,1,3]]$ code
introduced in Refs.~\cite{laflamme1996fivequbitcode} and \cite{divincenzo1996fivequbitcodeimplementation}.

\subsection{Distance-two surface code}

Surface codes are considered to be among the most promising error-correction
schemes as they can cope with error rates of about $10^{-2}$~\cite{Fowler2012surfacecode}.
The distance-two surface code $[[ 5,1,2]]$ studied in Ref.~\cite{DEVI16}
employs postselection to discard outcomes which are not included in the code space.
In other words, the set of
5-bit strings resulting from the $N=8192$ shots are first analyzed and then the results are corrected.

\begin{figure}[ht]
\begin{center}
\includegraphics[width=0.8\hsize]{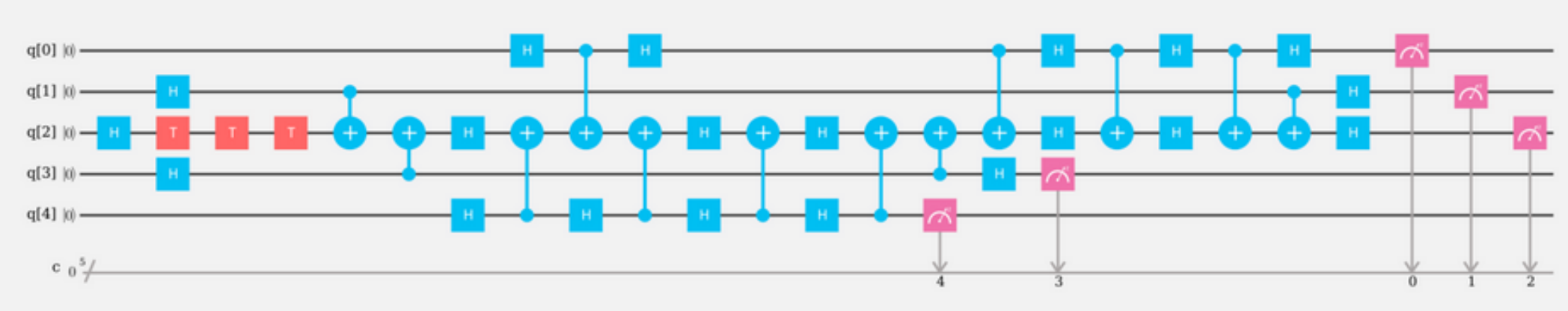}
\caption{(color online) %
Circuit with three $T$-gates performing a rotation of the third qubit $Q_2$, followed by the encoding circuit
of a distance-two surface code.
}
\label{fig5a}
\end{center}
\end{figure}

\begin{figure}[ht]
\begin{center}
\includegraphics[height=0.15\hsize]{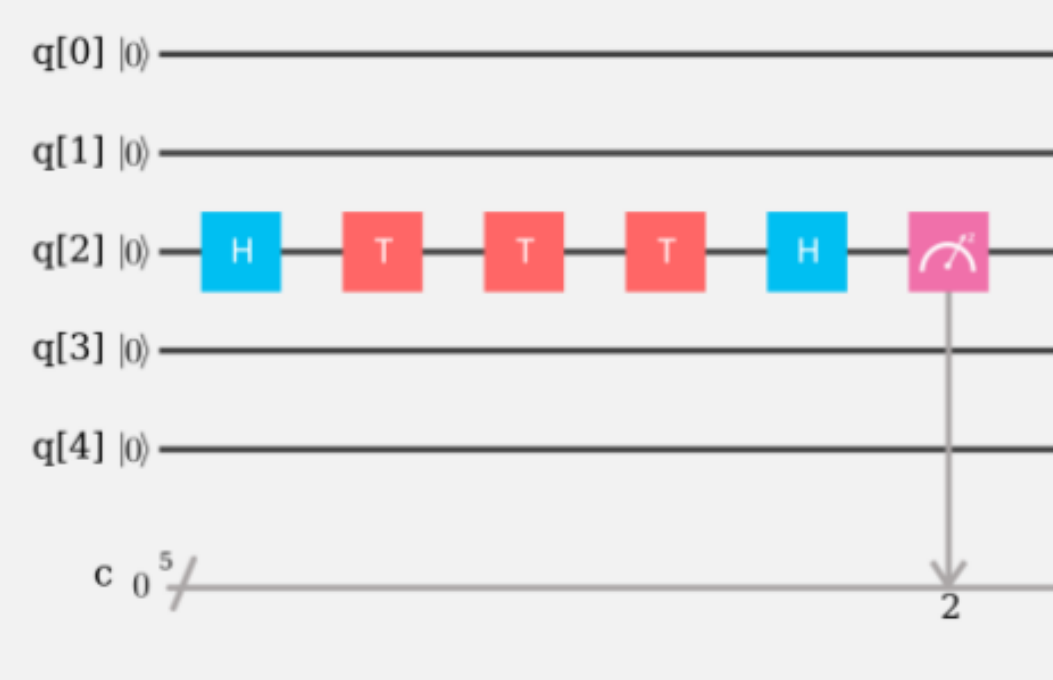}
\caption{(color online) %
Circuit with three $T$-gates performing a rotation of a single qubit.
}
\label{fig5b}
\end{center}
\end{figure}

In our first test of the distance-two surface code, we employ the same encoding circuit as described by Devitt~\cite{DEVI16}.
This circuit is shown in Fig.~\ref{fig5a} and starts right after the last of the three $T$-gates.
In our test procedure, the chosen number of $T$-gates ranges from 0 to 8, eventually resulting in
a rotation of the state $|0\rangle$ to $|1\rangle$ and back to $|0\rangle$.
For comparison, we also perform the same test without quantum error correction.
As an example, we show the corresponding circuit with three $T$-gates in Fig.~\ref{fig5b}.

In our second test, we use the same encoding circuit as in the first test, followed by $K=0,\ldots,8$ logical
$X$-gates.
In contrast to our first test, the state $|0\rangle$ is first encoded into the logical state $|0\rangle_L$
and then the logical-$X$ operator is applied $K$ times before the measurements in the $z$-basis are performed.
For comparison, the same experiment is repeated using a single qubit without the quantum error correction.

Postselection is done according to Table~\ref{tab5a}, that is we only count the outcomes that correspond to a measurement of one of
the logical states $|0\rangle_L $ or $|1\rangle_L$.
In our experiments, about three quarters of the shots are discarded by the postselection process.

\setcounter{table}{3}
\begin{table}[ht]
\begin{center}
\caption{List of measured 5-bit strings that correspond to the logical states $|0\rangle_L$ or $|1\rangle_L$.
Outcomes that correspond to neither of the logical states are omitted.
Note that the order of the qubits differs from that chosen in Ref.~\cite{DEVI16}.
}
\begin{tabular}{cc}
\hline\smallskip
$|0\rangle_L$ &  $|1\rangle_L$ \\
\hline
$00000$ & $00011$ \\
$01111$ & $01100$ \\
$10110$ & $10101$ \\
$11001$ & $11010$ \\
\hline
\end{tabular}
\label{tab5a}
\end{center}
\end{table}

\begin{figure}[ht]
\begin{center}
\includegraphics[width=0.49\hsize]{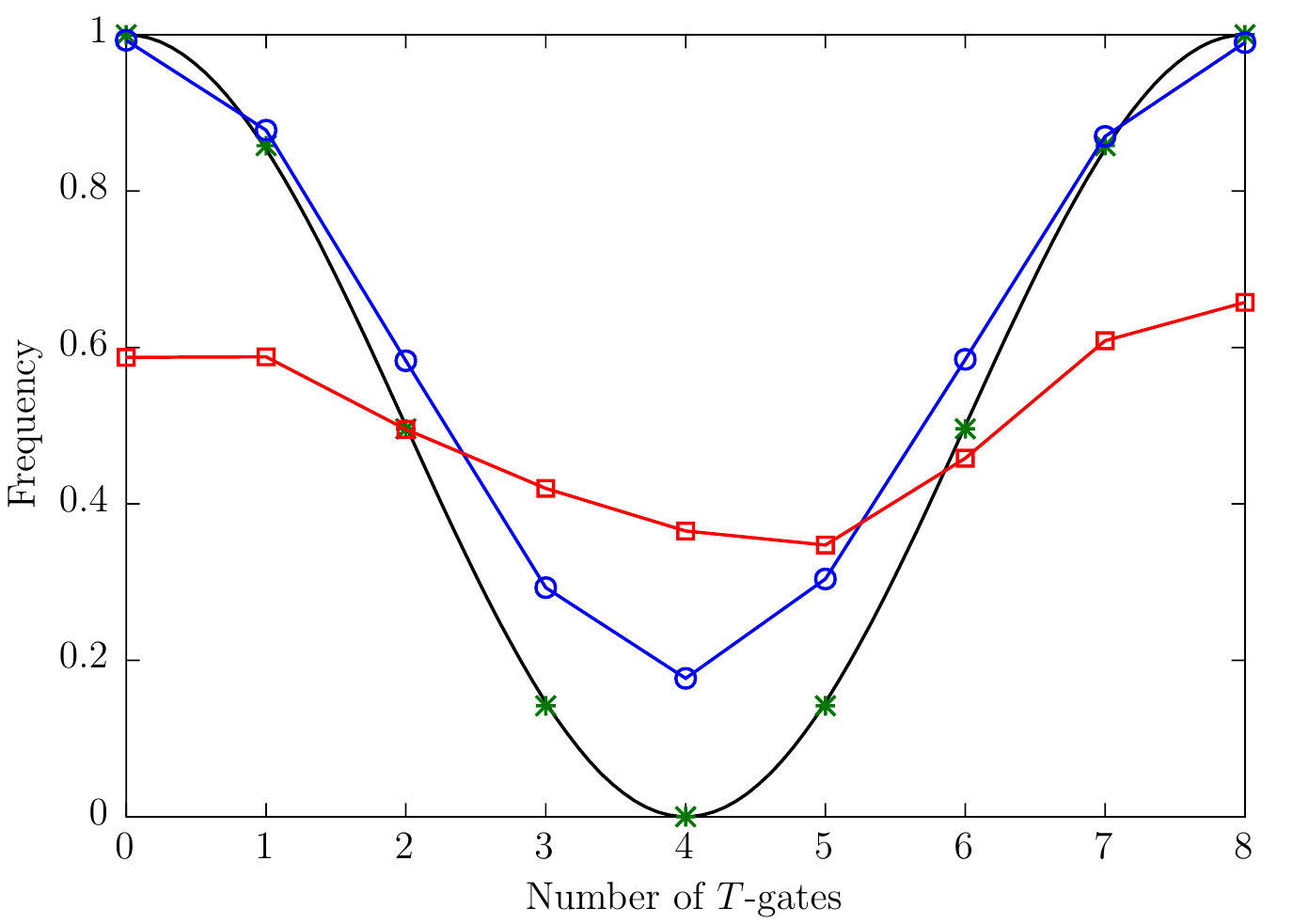}
\includegraphics[width=0.49\hsize]{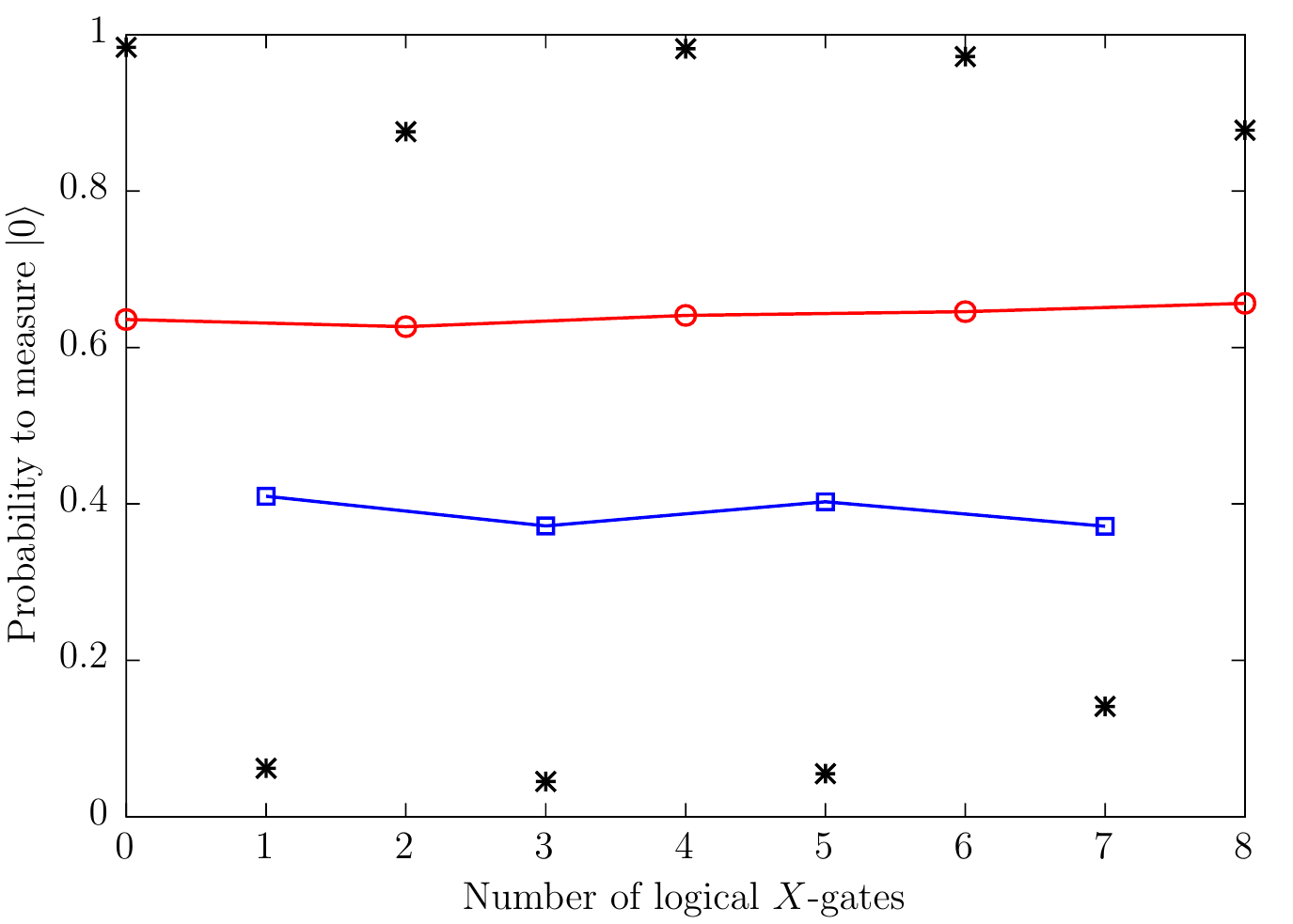}
\caption{(color online) %
Left: first experiment with the distance-two surface code $[[ 5,1,2]]$, previously also performed by Devitt~\cite{DEVI16}.
Solid line (black): prediction of quantum theory ($\cos^2(\pi K/8)$) for the ideal quantum computer,
$K$ denoting the number of $T$-gates;
stars (green): data generated by the simulator;
open circles (blue): single qubit circuit, see Fig.~\ref{fig5b};
open squares (red): results obtained by postselection of the data produced by the quantum error-correction circuit shown in Fig.~\ref{fig5a}.
Right: results of the second experiment with the same code. For an even (odd) number of $X$-gates, quantum theory predicts
the probability to measure $|0\rangle$ to be 1 (0).
Stars (black): single qubit circuit;
open circles (red): quantum error-corrected results for an even number of $X$ gates;
open squares (blue): quantum error-corrected results for an odd number of $X$ gates.
All experiments have been carried out in January 2017.
Lines connecting the data points are guides to the eye.
}
\label{fig5c}
\end{center}
\end{figure}

The results of the first test are shown in Fig.~\ref{fig5c} (left). Stars correspond to the output of the simulator which operates
as an ideal quantum computer and therefore provides a stringent test of the correctness of the circuit itself.
The outcomes lie on the solid line which represents the sinusoidal function predicted by quantum theory.
The results from the single-qubit circuit (open circles) qualitatively follow the expected curve
but with a reduced visibility.
For the encoded qubit, postselection reduces the number of valid shots from 8192 to 2000 -- 2300, i.e.,
there are about 2000 5-bit strings that correspond either to a logical 0 or a logical 1.
After postselection, the results of the encoded qubit (open squares) are worse than for the single qubit,
with deviations from the exact result that are far outside the range of the statistical fluctuations.
However, a potential problem of this first test is that the rotation is done on the single qubit before it is encoded.
Hence errors that occur during the rotation cannot be detected by the error-correction code.

In our second test, see Fig.~\ref{fig5c} (right), the rotation is thus performed after the encoding but the results are essentially
the same: the encoding procedure induces more errors than can be detected.
Indeed, the single qubit results (stars) are much closer to the ideal outcome (one for
an even number of $X$-gates and zero for an odd number of $X$-gates) when no error-correcting code is used
than those obtained with the use of error-correction.
The results for the encoded qubit (open circles for an even number of $X$-gates and open squares for an odd
number) are only slightly better than just randomly picking zeros or ones.
At least, the frequencies of the quantum error-corrected qubits do not change a lot with the number of $X$-gates applied.

\subsection{Distance-three 5-qubit code}

\begin{figure}[t]
\begin{center}
\includegraphics[trim={0 15cm 0 0},width=\hsize,clip]{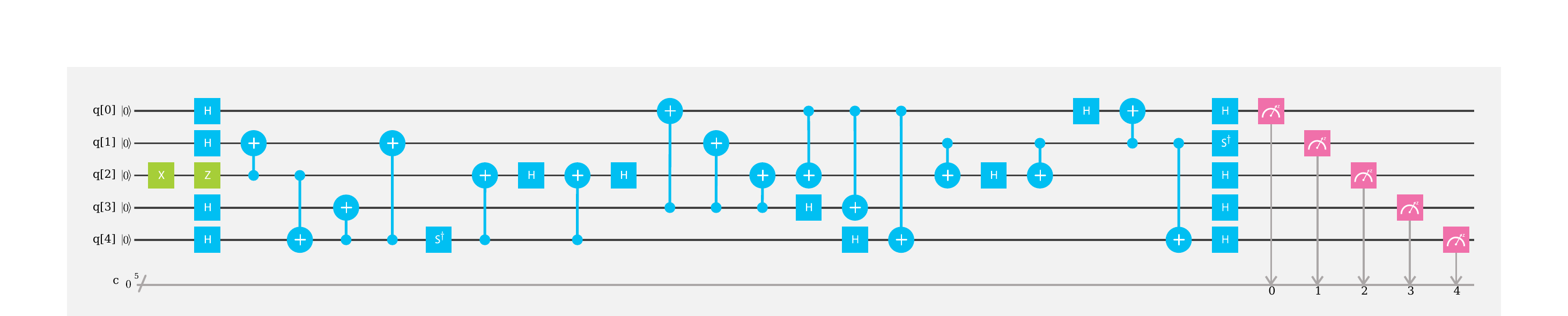}
\caption{(color online) %
Circuit diagram to encode the central qubit $|Q_2\rangle$ into the logical codeword $|Q_2\rangle_L$ given by
Eqs.~(\ref{logical0}) and (\ref{logical1}). The diagram has been taken from Ref.~\cite{niwa2002simulatingQECC} and adapted to the
set of gates that the IBM-QE can execute. The full circuit, obtained by re-expressing
the CNOT gates that cannot be executed by the IBM-QE hardware, is given in \ref{appA}.
}
\label{fig5e}
\end{center}
\end{figure}

The distance-three 5-qubit code $[[ 5,1,3 ]]$ is called a \emph{perfect} quantum error-correcting code since it is the smallest code that, theoretically, has
the ability to correct any single-qubit error \cite{laflamme1996fivequbitcode}. From the different presentations of
this code (see Ref.~\cite{divincenzo1996fivequbitcodeimplementation}), we choose the one given in Refs.~\cite{NIEL10} and
\cite{niwa2002simulatingQECC} where the logical states are defined as

\begin{eqnarray}
  |0\rangle_L &=& \frac 1 4 (+|00000\rangle - |00011\rangle + |00101\rangle - |00110\rangle + |01001\rangle + |01010\rangle \nonumber\\
  &&\ \,\ \ \ - |01100\rangle - |01111\rangle - |10001\rangle + |10010\rangle + |10100\rangle - |10111\rangle \nonumber \\
  &&\ \,\ \ \ - |11000\rangle - |11011\rangle - |11101\rangle - |11110\rangle)
  \label{logical0},
  \\
  |1\rangle_L &=& \frac 1 4 (-|00001\rangle - |00010\rangle - |00100\rangle - |00111\rangle - |01000\rangle + |01011\rangle \nonumber \\
  &&\ \,\ \ \  + |01101\rangle - |01110\rangle - |10000\rangle - |10011\rangle + |10101\rangle + |10110\rangle \nonumber \\
  &&\ \,\ \ \ - |11001\rangle + |11010\rangle - |11100\rangle + |11111\rangle)
  \label{logical1} .
\end{eqnarray}
The encoding circuit for this code is given in Ref.~\cite{niwa2002simulatingQECC}.
As the IBM-QE does not support the controlled $Z$ ($CZ$), controlled $-Z$ ($C\!-\!Z$),
and controlled $Y$ ($CY$) gates,
we rewrite the circuit in terms of the CNOT gate ($C$) and other single-qubit gates
that the IBM-QE supports by using the circuit identities
\begin{eqnarray}
  CZ_{ij} &=& H_j C_{ij} H_j \label{controlledZ},\\
  C\!-\!Z_{ij} &=& H_j C_{ij} H_j Z_i \label{controlled-Z},\\
  CY_{ij} &=& H_j C_{ij} H_j C_{ij} S_i \label{controlledY}.
\end{eqnarray}
Applying the relation $SZ=S^\dagger$ and swapping the lines of $Q_0$ and $Q_2$, we thus arrive at the encoding circuit shown in
Fig.~\ref{fig5e}, whose purpose is to encode the state $|00Q_200\rangle\mapsto|Q_2\rangle_L$. Due to the reduced connectivity of
the IBM-QE, we further express all unsupported CNOT gates $C_{ij}$ in terms of $C_{i2}$ and $C_{j2}$ using the identity
$C_{ij}=H_iH_jC_{ji}H_iH_j$ and the SWAP gate $\mathrm{SWAP}_{ij}=C_{ij}C_{ji}C_{ij}$. The listing of the full circuit is given in
\ref{appA} and takes about $33\;\mu\mathrm{s}$ to run to completion.

\begin{figure}[ht]
\begin{center}
\includegraphics[width=0.49\hsize]{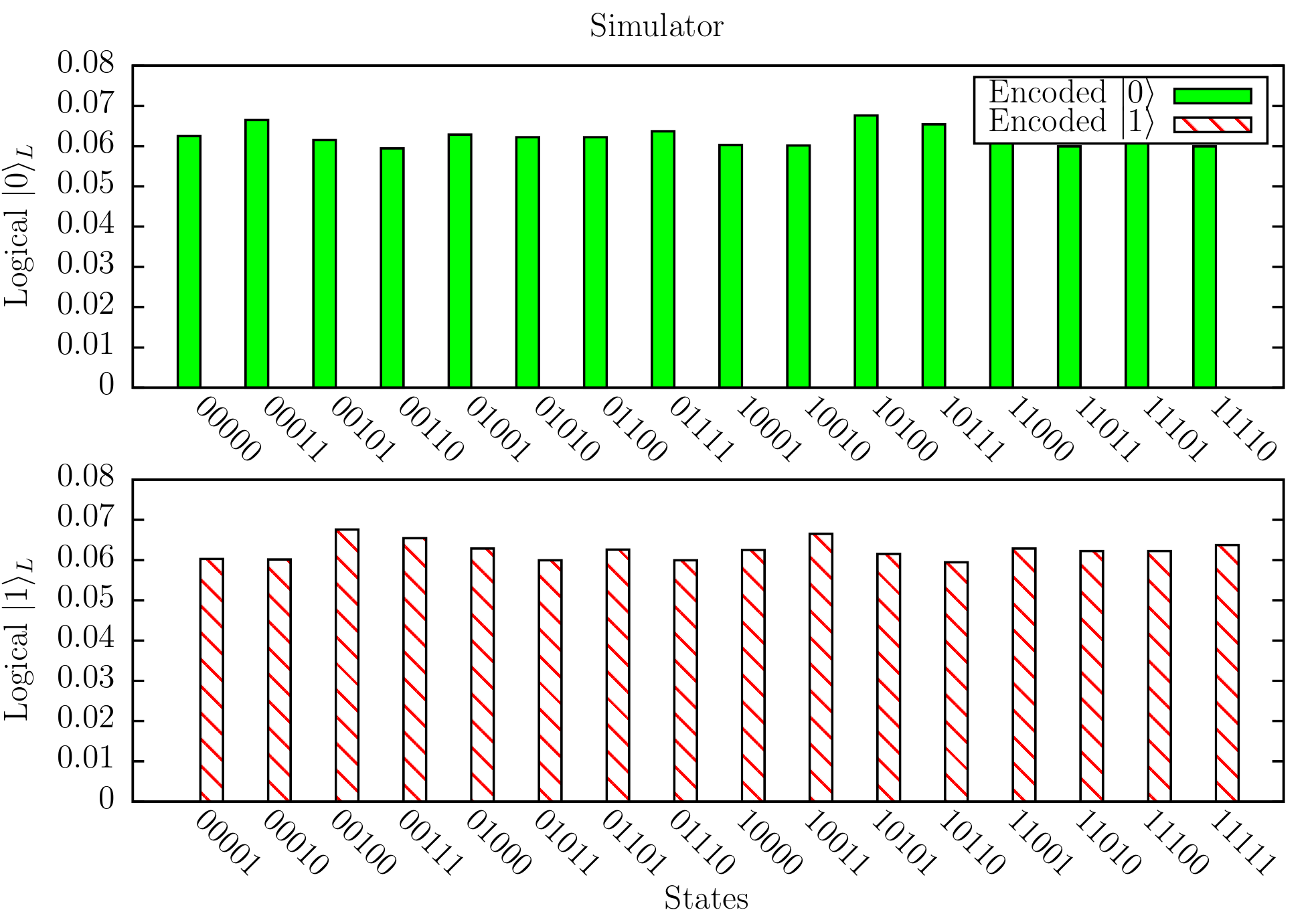}
\includegraphics[width=0.49\hsize]{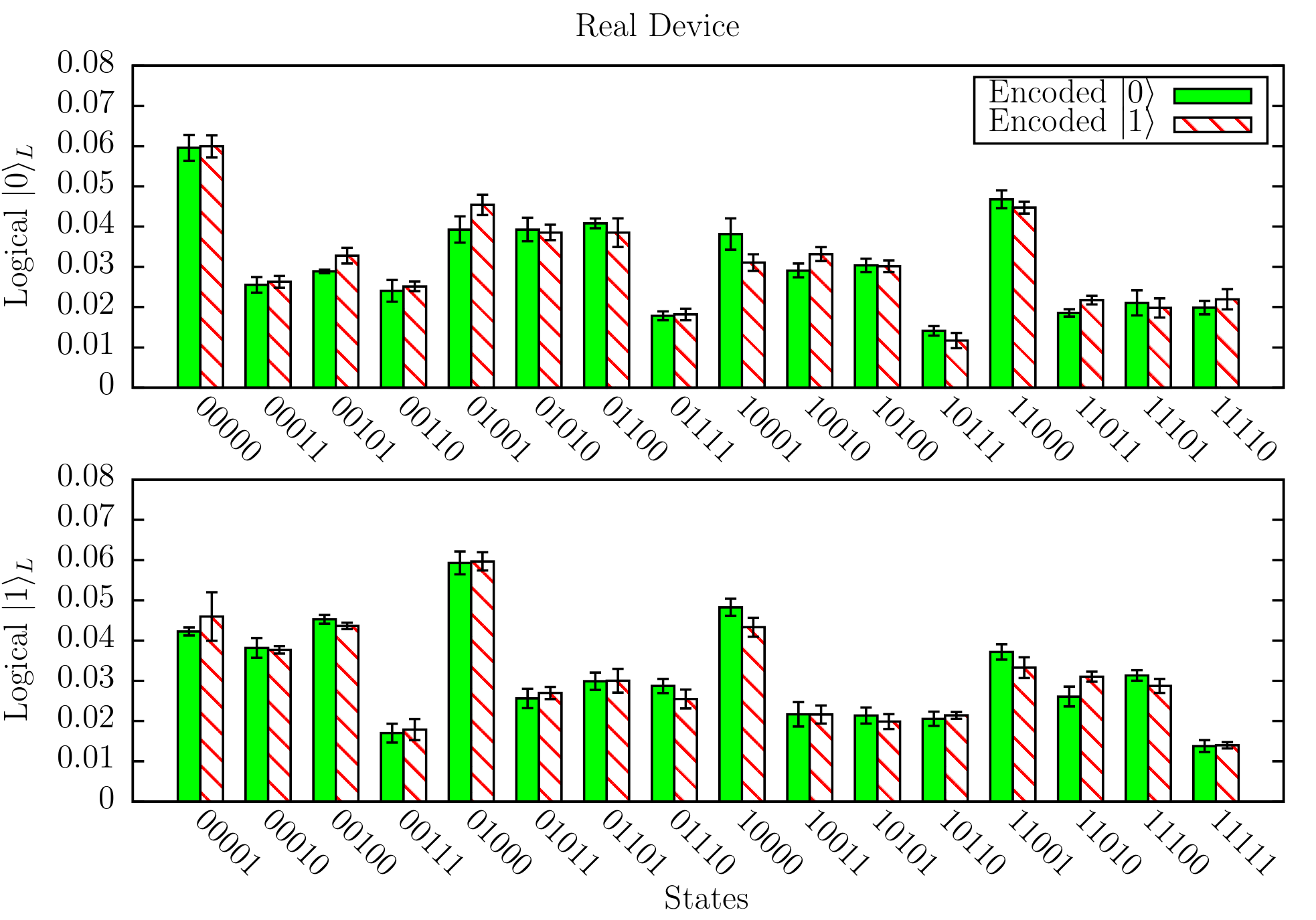}
\caption{(color online) %
Results of the encoding part of the perfect distance-three 5-qubit code.
Left: results obtained by the simulator, i.e. by simulating the ideal quantum computer.
Right: results produced by the IBM-QE processor on January 26, 2017.
Shown are the resulting frequencies of all the 32 basis states,
grouped into those constituting $|0\rangle_L$ (top row) and those constituting $|1\rangle_L$ (bottom row).
Solid (hatched) bars correspond to the initial state $|00Q_200\rangle=|00000\rangle$ ($|00Q_200\rangle=|00100\rangle$).
The standard deviations resulting from five independent runs on the hardware processor (each with $N=8192$ shots)
are shown as error bars.
}
\label{fig5f}
\end{center}
\end{figure}

The correctness of the encoding circuit is established by running the ciruit on the ideal quantum computer simulator
(included in the IBM-QE).
The results are shown in the left panel of Fig.~\ref{fig5f}.
From the top-left panel, it is clear that starting from the state $|Q_2=0\rangle$,
the circuit produces a uniform superposition of all the basis states (solid bars) contained in the codeword $|0\rangle_L$ (see Eq.~(\ref{logical0})).
Similarly, the bottom-left panel shows that encoding the state $|Q_2=1\rangle$ properly yields the codeword $|1\rangle_L$
(hatched bars) given by Eq.~(\ref{logical1}).
Note that the height of the bars differs slightly from $1/16$ due to the random sampling used in the $N=8192$ shots.

The results of running the same circuit on the real chip are depicted in the right panel of Fig.~\ref{fig5f}.
We repeated each experiment of $N=8192$ shots five times to get some information about the reproducibility
and the statistical distribution of the results.
In the right panel of Fig.~\ref{fig5f}, the corresponding standard deviations are indicated by error bars.

The results clearly demonstrate that, apart from statistical fluctuations,
the real processor produces the same output irrespective of whether the initial state
was $|Q_2=0\rangle$ (solid bars) or $|Q_2=1\rangle$ (hatched bars).
The outcome is completely different from the one obtained with the simulator (compare with left panel of Fig.~\ref{fig5f}).
For each case, the resulting distribution contains in a pair-wise manner almost equally important contributions
from the other case.
This makes it impossible for the user to distinguish the logical state $|0\rangle_L$
from the logical state $|1\rangle_L$.
Therefore, we have to conclude that the IBM-QE fails to generate the correct outcome
for the encoding part of the perfect distance-three 5-qubit code.
Unfortunately, because the number of gates that is allowed on
the IBM-QE is limited to 80, we cannot carry out the corresponding decoding circuit.

\section{Discussion}\label{section6}

We have explored the use of four classes of quantum circuits to benchmark
quantum computer hardware by executing these circuits on the only gate-based quantum
computer that is publicly accessible today.
The class of identity operations built from CNOT gates stands out
in terms of simplicity, scalability, and sensitivity to malfunctioning hardware.
We propose that apart from characterizing the operation of the individual qubits,
and as a minimal benchmark, any system that performs quantum computation is subjected
to this class of circuits.
Some of these circuits might also be useful for the calibration procedure itself.

From the results presented in the foregoing sections, we draw the following conclusions:
\begin{itemize}
\item
For some systems of two and four qubits, qualitative agreement with quantum theory was observed.
\item
Errors could not be identified by the user nor be corrected using quantum error-correction,
and could not be attributed to the specified gate errors.
\item
The data showed strong variations between calibrations.
\item
Sequences of identity operations provide simple, scalable algorithms to validate the correct operation of the device~\cite{RAED02}.
\item
The current IBM-QE device does not meet the two elementary requirements (see section~\ref{section0}) for a computing device.
\item
The IBM-QE allows a theoretician to perform real laboratory experiments.
\end{itemize}

From the perspective of a user, the IBM-QE does not perform as could reasonably be expected from a computer.
Except for very simple circuits which return qualitatively correct results,
the IBM-QE device often fails to return the correct results for reasons which in some cases
may be traced back to running the algorithm on a different day (with a different device calibration)
but in other cases do not seem to have a simple explanation.
Needless to say, it would be of great interest to have the simple benchmarks
carried out on other hardware platforms, in particular on the recent 5-qubit ion-trap device~\cite{MONR17}, and
see how they perform relative to the IBM-QE.

One fairly simple reason for the failure of the IBM-QE to function as a computer
may be that the two-state model used to describe the qubits does not capture,
not even approximately, the time evolution of the system of coupled transmons~\cite{KOCH07}.
Indeed, preliminary simulations based on a more comprehensive model of transmons
indicate that their time evolution fundamentally involves more than two energy levels.
This then raises the question whether the failures observed in our IBM-QE experiments
can be traced back to the limited usefulness of the two-state description.
This question can readily be addressed by solving the time-dependent Schr\"odinger equation
for more realistic models of coupled transmons and we intend to carry out such simulations in the near future.

\section*{Acknowledgements}
We are grateful to the IBM Quantum Experience project team for sharing technical details with us.
This work does not reflect the views or opinions of IBM or any of its employees.
\clearpage
\setcounter{table}{0}

\begin{sidewaystable}[ht]
\caption{
Data of IBM-QE experiments with circuits which add two 2-bit numbers modulo 4 for the case that each pair of qubits
encodes a single integer in the range 0 to 3.
The first (second) group of results has been obtained with a circuit that uses qubits 0 -- 3 (1 -- 4).
See \ref{appA} for more details on the circuits used.
The experimental outcomes are colored according to the rule
green: correct;
red: wrong;
magenta: unexpected (wrong) superposition;
black: states with the next-to-highest relative frequency.
The numbers in parentheses are the relative frequencies of occurrence.
}
\begin{tabularx}{\textwidth}{@{\extracolsep{\fill}} ccccc}
\noalign{\medskip}\hline
Operation & \hfil Outcome \hfil & \multicolumn{2}{c}{Outcome} &\hfil Date \hfil\\
\hline\hline\noalign{\smallskip}
$|Q_0 Q_1\rangle+|Q_2 Q_3\rangle\rightarrow$ &           Quantum Theory            & \multicolumn{2}{c}{IBM-QE (8192 shots)}\\
$|Q_3 Q_2 Q_0 Q_1\rangle$ &$|\mathrm{state}\rangle$\ \ (prob.) &  $|\mathrm{state}\rangle$\ \ (frequency)& $|\mathrm{state}\rangle$\ \ (frequency)\\
\hline\noalign{\smallskip}
$2+1=3$ & $\STATE{1}{1}{1}{0}$\ \ (1.00)  & {\color{red}$\STATE{1}{1}{0}{1}$\ \ (0.275)}& {\color{black}$\STATE{0}{0}{0}{1}$\ \ (0.160)}& 22-Jan-17\\
$2+3=1$ & $\STATE{0}{1}{0}{1}$\ \ (1.00)  & {\color{DarkGreen}$\STATE{0}{1}{0}{1}$\ \ (0.318)}& {\color{black}$\STATE{0}{1}{0}{0}$\ \ (0.150)}& 23-Jan-17\\
 &  & {\color{red}$\STATE{0}{0}{0}{1}$\ \ (0.271)}& {\color{black}$\STATE{1}{0}{0}{1}$\ \ (0.169)}& 30-Jan-17\\
 &  & {\color{red}$\STATE{0}{0}{0}{1}$\ \ (0.277)}& {\color{black}$\STATE{1}{0}{0}{1}$\ \ (0.172)}& 30-Jan-17\\
$1+3=0$ & $\STATE{0}{0}{1}{0}$\ \ (1.00)  & {\color{DarkGreen}$\STATE{0}{0}{1}{0}$\ \ (0.343)}& {\color{black}$\STATE{0}{0}{0}{0}$\ \ (0.221)}& 30-Jan-17\\
\hline\hline\noalign{\smallskip}
$|Q_1 Q_3\rangle+|Q_2 Q_4\rangle\rightarrow$\\
$|Q_2 Q_4 Q_1 Q_3\rangle$\\
\hline
$2+1=3$ & $\ASTATE{1}{0}{1}{1}$\ \ (1.00)  & {\color{magenta}$\ASTATE{1}{0}{0}{1}$\ \ (0.342)}& {\color{magenta}$\ASTATE{1}{0}{1}{1}$\ \ (0.341)}& 25-Jan-17\\
        & & {\color{DarkGreen}$\ASTATE{1}{0}{1}{1}$\ \ (0.315)}& {\color{black}$\ASTATE{0}{0}{0}{1}$\ \ (0.165)}& 30-Jan-17\\
$2+3=1$ & $\ASTATE{1}{0}{0}{1}$\ \ (1.00)  & {\color{red}$\ASTATE{1}{0}{1}{1}$\ \ (0.225)}& {\color{black}$\ASTATE{0}{0}{0}{1}$\ \ (0.199)}& 30-Jan-17\\
        & & {\color{red}$\ASTATE{0}{0}{0}{1}$\ \ (0.244)}& {\color{black}$\ASTATE{1}{0}{1}{1}$\ \ (0.210)}& 30-Jan-17\\
$1+3=0$ & $\ASTATE{1}{0}{0}{0}$\ \ (1.00)  & {\color{DarkGreen}$\ASTATE{1}{0}{0}{0}$\ \ (0.493)}& {\color{black}$\ASTATE{1}{1}{0}{0}$\ \ (0.154)}& 30-Jan-17\\
\hline
\label{tab2a}
\end{tabularx}
\end{sidewaystable}

\begin{sidewaystable}[ht]
\caption{
Data of IBM-QE experiments with circuits which add two 2-bit numbers modulo 4
for the case in which the inputs are superpositions of two or four states.
See \ref{appA} for an example of the circuit used.
In all cases shown the IBM-QE yields the correct outcomes.
}
\begin{tabularx}{\textwidth}{@{\extracolsep{\fill}} ccccc}
\noalign{\medskip}\hline
Operation & \hfil Outcome \hfil  & \hfil Outcome \hfil &Date \hfil\\
\hline\hline\noalign{\smallskip}
$|Q_0 Q_1\rangle+|Q_2 Q_3\rangle\rightarrow|Q_3 Q_2 Q_0 Q_1\rangle$ &Quantum Theory& IBM-QE (8192 shots)\\
Superposition of& $|\mathrm{state}\rangle$\ \ (prob.) &  $|\mathrm{state}\rangle$\ \ (frequency)\\
\hline
$0+0=0$ & $\STATE{0}{0}{0}{0}$\ \ (0.50)  & {\color{DarkGreen}$\STATE{0}{0}{0}{0}$\ \ (0.354)}& 17-Jan-17\\
$1+0=1$ & $\STATE{0}{1}{1}{0}$\ \ (0.50)  & {\color{DarkGreen}$\STATE{0}{1}{1}{0}$\ \ (0.311)}& 17-Jan-17\\
\hline
$1+0=1$ & $\STATE{0}{1}{1}{0}$\ \ (0.50)  & {\color{DarkGreen}$\STATE{0}{1}{1}{0}$\ \ (0.314)}& 17-Jan-17\\
$1+3=0$ & $\STATE{0}{0}{1}{0}$\ \ (0.50)  & {\color{DarkGreen}$\STATE{0}{0}{1}{0}$\ \ (0.261)}& 17-Jan-17\\
\hline
$1+0=1$ & $\STATE{0}{1}{1}{0}$\ \ (0.25)  & {\color{DarkGreen}$\STATE{0}{1}{1}{0}$\ \ (0.185)}& 17-Jan-17\\
$1+3=0$ & $\STATE{1}{1}{1}{1}$\ \ (0.25)  & {\color{DarkGreen}$\STATE{1}{1}{1}{1}$\ \ (0.152)}& 17-Jan-17\\
$3+0=3$ & $\STATE{0}{0}{1}{0}$\ \ (0.25)  & {\color{DarkGreen}$\STATE{0}{0}{1}{0}$\ \ (0.147)}& 17-Jan-17\\
$3+3=2$ & $\STATE{1}{0}{1}{0}$\ \ (0.25)  & {\color{DarkGreen}$\STATE{1}{0}{1}{0}$\ \ (0.097)}& 17-Jan-17\\
\hline
\label{tab2b}
\end{tabularx}
\end{sidewaystable}

\begin{sidewaystable}[ht]
\caption{
Data for experiments with identity operations, generated by the IBM-QE.
See \ref{appA} for more details on the circuits used.
Green (red)-colored states indicate experimental outcomes that, according to the rule (see section~\ref{section1}) used,
are (not) the same as the theoretically expected state.
The states in the experimental outcomes with the next-to-highest relative frequency are given in black.
The numbers in parentheses are the relative frequencies of occurrence.
Each calculation has been repeated several times, on the dates indicated.
}
\begin{tabularx}{\textwidth}{@{\extracolsep{\fill}} cccccc}
\noalign{\medskip}\hline
Operation & \hfil Input \hfil  & \hfil Outcome \hfil & \multicolumn{2}{c}{Outcome} &\hfil Date \hfil\\
\hline\hline\noalign{\smallskip}
&  &           Quantum Theory            & \multicolumn{2}{c}{IBM-QE (8192 shots)}\\
&  & $|\mathrm{state}\rangle$\ \ (prob.) &  $|\mathrm{state}\rangle$\ \ (frequency)& $|\mathrm{state}\rangle$\ \ (frequency)\\
\hline
$\left(C_{01}\right)^8$ & $\state{0}{1}{0}{0}$ & $\state{0}{1}{0}{0}$\ \ (1.00) & {\color{DarkGreen}$\state{0}{1}{0}{0}$\ \ (0.661)}& $\state{1}{1}{0}{0}$\ \ (0.299)&16-Jan-17\\
         & & & {\color{DarkGreen}$\state{0}{1}{0}{0}$\ \ (0.700)}& $\state{1}{1}{0}{0}$\ \ (0.198)&18-Jan-17\\
                                                                & & & {\color{DarkGreen}$\state{0}{1}{0}{0}$\ \ (0.642)}& $\state{1}{1}{0}{0}$\ \ (0.289)&19-Jan-17\\
                                                                & & & {\color{DarkGreen}$\state{0}{1}{0}{0}$\ \ (0.580)}& $\state{1}{1}{0}{0}$\ \ (0.335)&23-Jan-17\\
                                                                & & & {\color{DarkGreen}$\state{0}{1}{0}{0}$\ \ (0.628)}& $\state{1}{1}{0}{0}$\ \ (0.256)&23-Jan-17\\
                                                                \\
\hline
$\left(C_{34}\right)^8$ & $\state{0}{4}{0}{3}$ & $\state{0}{4}{0}{3}$\ \ (1.00) & {\color{red}$\state{1}{4}{0}{3}$\ \ (0.512)}& $\state{0}{4}{0}{3}$\ \ (0.372)&15-Jan-17\\
                    & &  & {\color{red}$\state{1}{4}{0}{3}$\ \ (0.567)}& $\state{0}{4}{0}{3}$\ \ (0.318)&16-Jan-17\\
                                                                           & &  & {\color{red}$\state{1}{4}{0}{3}$\ \ (0.548)}& $\state{0}{4}{0}{3}$\ \ (0.363)&18-Jan-17\\
                                                                           & &  & {\color{red}$\state{1}{4}{0}{3}$\ \ (0.616)}& $\state{0}{4}{0}{3}$\ \ (0.275)&19-Jan-17\\
                                                                           & &  & {\color{red}$\state{1}{4}{0}{3}$\ \ (0.590)}& $\state{0}{4}{0}{3}$\ \ (0.323)&22-Jan-17\\
                                                                           & &  & {\color{red}$\state{1}{4}{0}{3}$\ \ (0.618)}& $\state{0}{4}{0}{3}$\ \ (0.321)&23-Jan-17\\

\hline
$\left(C_{34}\right)^8$ & $\state{0}{4}{1}{3}$ & $\state{0}{4}{1}{3}$\ \ (1.00) & {\color{DarkGreen}$\state{0}{4}{1}{3}$\ \ (0.794)}& $\state{0}{4}{0}{3}$\ \ (0.084)&15-Jan-17\\
                    & &  & {\color{DarkGreen}$\state{0}{4}{1}{3}$\ \ (0.797)}& $\state{0}{4}{0}{3}$\ \ (0.088)&18-Jan-17\\
                                                                           & &  & {\color{DarkGreen}$\state{0}{4}{1}{3}$\ \ (0.853)}& $\state{0}{4}{0}{3}$\ \ (0.077)&23-Jan-17\\
                                                                           & &  & {\color{DarkGreen}$\state{0}{4}{1}{3}$\ \ (0.849)}& $\state{0}{4}{0}{3}$\ \ (0.068)&23-Jan-17\\
\\
\\
\hline
$(C_{02}C_{12})^2 (C_{02})^2(C_{12})^2 $& $\CSTATE{1}{1}{1}$ & $\CSTATE{1}{1}{1}$\ \ (1.00) & {\color{DarkGreen}$\CSTATE{1}{1}{1}$\ \ (0.355)}& {\color{black}$\CSTATE{1}{1}{0}$\ \ (0.304)}&10-Jan-17\\
$(C_{02}C_{12})^2$ & &  & {\color{red}$\CSTATE{0}{1}{1}$\ \ (0.262)}& $\CSTATE{1}{1}{1}$\ \ (0.238)&18-Jan-17\\
                                                           & &  & {\color{red}$\CSTATE{0}{1}{1}$\ \ (0.250)}& $\CSTATE{1}{1}{1}$\ \ (0.237)&18-Jan-17\\
                                                           & &  & {\color{red}$\CSTATE{0}{1}{1}$\ \ (0.358)}& $\CSTATE{1}{1}{1}$\ \ (0.131)&27-Jan-17\\
                                                           & &  & {\color{red}$\CSTATE{0}{1}{1}$\ \ (0.368)}& $\CSTATE{1}{1}{1}$\ \ (0.128)&27-Jan-17\\
                                                           & &  & {\color{red}$\CSTATE{0}{1}{1}$\ \ (0.347)}& $\CSTATE{1}{1}{1}$\ \ (0.139)&27-Jan-17\\
                                                           & &  & {\color{red}$\CSTATE{0}{1}{1}$\ \ (0.374)}& $\CSTATE{1}{1}{1}$\ \ (0.150)&27-Jan-17\\
\hline
$H_0H_1X_0X_1(C_{02}C_{12})^2 $ &  $\CSTATE{1}{1}{1}$ & $\CSTATE{1}{1}{1}$\ \ (1.00) & {\color{DarkGreen}$\CSTATE{1}{1}{1}$\ \ (0.304)}& {\color{black}$\CSTATE{0}{1}{1}$}\ \ (0.157)&05-May-16\\
$(C_{02})^2(C_{12})^2 $& &  & {\color{DarkGreen}$\CSTATE{1}{1}{1}$\ \ (0.298)}& $\CSTATE{0}{1}{1}$\ \ (0.192)&09-Nov-16\\
$(C_{02}C_{12})^2H_0H_1$  & &  & {\color{DarkGreen}$\CSTATE{1}{1}{1}$\ \ (0.223)}& $\CSTATE{0}{1}{1}$\ \ (0.156)&20-Nov-16\\
  & &  & {\color{red}$\CSTATE{0}{1}{1}$\ \ (0.232)}& $\CSTATE{0}{0}{0}$\ \ (0.154)&27-Jan-17\\
  & &  & {\color{red}$\CSTATE{0}{1}{1}$\ \ (0.202)}& $\CSTATE{0}{0}{0}$\ \ (0.170)&27-Jan-17\\
\hline
\label{tab1}
\end{tabularx}
\end{sidewaystable}

\clearpage
\appendix
\section{Algorithms in QASM language}\label{appA}
For completeness, we give the .qasm files of the quantum-gate circuits used to perform
the experiments reported on in this paper.
For a detailed description of the programming language see Ref.~\cite{IBMQE}.

\subsection{Singlet state}
\begin{lstlisting}[language=Fortran]
IBMQASM 2.0;

include "qelib1.inc";

qreg q[5];
creg c[5];

x q[1];
x q[2];
h q[1];
cx q[1],q[2];
h q[1];
h q[2];
u1(pi/180*0) q[1];
u1(pi/180*0) q[2];
h q[1];
h q[2];
measure q[1] -> c[1];
measure q[2] -> c[2];
\end{lstlisting}

\subsection{Adder using qubits 0 -- 3}
\begin{lstlisting}[language=Fortran]
IBMQASM 2.0;
include "qelib1.inc";

qreg q[5];
creg c[5];

x q[1];
x q[2];
x q[3];

h q[2];
t q[2];
cx q[3],q[2];
tdg q[2];
t q[3];
cx q[3],q[2];
h q[3];
s q[2];
cx q[0],q[2];
s q[0];
sdg q[2];
cx q[0],q[2];
t q[2];
cx q[1],q[2];
t q[1];
tdg q[2];
cx q[1],q[2];
cx q[3],q[2];
h q[2];
h q[3];
cx q[3],q[2];
h q[2];
h q[3];
cx q[3],q[2];
s q[2];
cx q[1],q[2];
s q[1];
sdg q[2];
cx q[1],q[2];
id q[3];
tdg q[3];
h q[3];
cx q[3],q[2];
h q[2];
h q[3];
tdg q[2];
t q[3];
h q[2];
h q[3];
cx q[3],q[2];
h q[2];
id q[3];
measure q[3] -> c[3];
measure q[2] -> c[2];
measure q[1] -> c[1];
measure q[0] -> c[0];
\end{lstlisting}
\subsection{Adder using qubits 1 -- 4}
\begin{lstlisting}[language=Fortran]
IBMQASM 2.0;
include "qelib1.inc";

qreg q[5];
creg c[5];

x q[2];
x q[3];
barrier q[0],q[1],q[2],q[3],q[4];
h q[2];
t q[2];
cx q[4],q[2];
tdg q[2];
t q[4];
cx q[4],q[2];
h q[4];
s q[2];
cx q[1],q[2];
s q[1];
sdg q[2];
cx q[1],q[2];
t q[2];
cx q[3],q[2];
tdg q[2];
t q[3];
cx q[3],q[2];
id q[4];
s q[4];
cx q[3],q[4];
s q[3];
sdg q[4];
cx q[3],q[4];
h q[4];
tdg q[2];
cx q[4],q[2];
t q[2];
tdg q[4];
cx q[4],q[2];
h q[2];
measure q[1] -> c[1];
measure q[2] -> c[2];
measure q[3] -> c[3];
measure q[4] -> c[4];
\end{lstlisting}

\subsection{Identity operation}
\begin{lstlisting}[language=Fortran]
IBMQASM 2.0;
include "qelib1.inc";

qreg q[5];
creg c[5];

cx q[0],q[1];
cx q[0],q[1];
cx q[0],q[1];
cx q[0],q[1];
cx q[0],q[1];
cx q[0],q[1];
cx q[0],q[1];
cx q[0],q[1];
measure q[0] -> c[0];
measure q[1] -> c[1];
\end{lstlisting}

\subsection{Error correction: distance-two surface code}
\begin{lstlisting}[language=Fortran]
IBMQASM 2.0;
include "qelib1.inc";

qreg q[5];
creg c[5];

h q[2];
t q[2];
t q[2];
t q[2];
h q[2];
measure q[2] -> c[2];
\end{lstlisting}
\begin{lstlisting}[language=Fortran]
IBMQASM 2.0;
include "qelib1.inc";

qreg q[5];
creg c[5];

h q[2];
h q[1];
t q[2];
h q[3];
t q[2];
t q[2];
cx q[1],q[2];
cx q[3],q[2];
h q[2];
h q[4];
h q[0];
cx q[4],q[2];
cx q[0],q[2];
h q[4];
h q[0];
cx q[4],q[2];
h q[2];
h q[4];
cx q[4],q[2];
h q[2];
h q[4];
cx q[4],q[2];
cx q[3],q[2];
measure q[4] -> c[4];
cx q[0],q[2];
h q[3];
h q[0];
h q[2];
measure q[3] -> c[3];
cx q[0],q[2];
h q[0];
h q[2];
cx q[0],q[2];
h q[0];
cx q[1],q[2];
h q[1];
h q[2];
measure q[0] -> c[0];
measure q[1] -> c[1];
measure q[2] -> c[2];
\end{lstlisting}
\begin{lstlisting}[language=Fortran]
IBMQASM 2.0;
include "qelib1.inc";

qreg q[5];
creg c[5];

x q[0]; // repeat 0-8 times
measure q[0] -> c[0];
\end{lstlisting}
\begin{lstlisting}[language=Fortran]
IBMQASM 2.0;
include "qelib1.inc";

qreg q[5];
creg c[5];

h q[1];
h q[2];
h q[3];
cx q[1],q[2];
cx q[3],q[2];
h q[2];
h q[4];
h q[0];
cx q[4],q[2];
cx q[0],q[2];
h q[4];
h q[0];
cx q[4],q[2];
h q[2];
h q[4];
cx q[4],q[2];
h q[2];
h q[4];
cx q[4],q[2];
cx q[3],q[2];
h q[4];
cx q[0],q[2];
h q[0];
h q[2];
cx q[0],q[2];
h q[0];
h q[2];
cx q[0],q[2];
cx q[1],q[2];

// repeat these logical x gates 0-8 times
z q[0];
z q[1];

h q[4];
h q[0];
h q[1];
h q[2];
h q[3];

measure q[4] -> c[4];
measure q[3] -> c[3];
measure q[0] -> c[0];
measure q[1] -> c[1];
measure q[2] -> c[2];
\end{lstlisting}
\subsection{Error correction: distance-three 5-qubit code}
\begin{lstlisting}[language=Fortran]
IBMQASM 2.0;
include "qelib1.inc";

qreg q[5];
creg c[5];

h q[0];
h q[1];
id q[2];
h q[3];
h q[4];
cx q[1],q[2];
h q[1];
h q[2];
cx q[1],q[2];
h q[1];
h q[2];
cx q[1],q[2];
cx q[4],q[2];
cx q[1],q[2];
h q[1];
h q[2];
cx q[1],q[2];
h q[1];
h q[2];
cx q[1],q[2];
sdg q[4];
cx q[4],q[2];
h q[2];
cx q[4],q[2];
h q[2];
cx q[0],q[2];
h q[0];
h q[2];
cx q[0],q[2];
h q[0];
h q[2];
cx q[0],q[2];
cx q[3],q[2];
cx q[0],q[2];
h q[0];
h q[2];
cx q[0],q[2];
h q[0];
h q[2];
cx q[0],q[2];
cx q[1],q[2];
h q[1];
h q[2];
cx q[1],q[2];
h q[1];
h q[2];
cx q[1],q[2];
cx q[3],q[2];
cx q[1],q[2];
h q[1];
h q[2];
cx q[1],q[2];
h q[1];
h q[2];
cx q[1],q[2];
cx q[3],q[2];
cx q[0],q[2];
h q[3];
h q[4];
cx q[3],q[2];
h q[2];
h q[3];
cx q[3],q[2];
h q[2];
h q[3];
cx q[3],q[2];
cx q[0],q[2];
cx q[3],q[2];
h q[2];
h q[3];
cx q[3],q[2];
h q[2];
h q[3];
cx q[3],q[2];
cx q[4],q[2];
h q[2];
h q[4];
cx q[4],q[2];
h q[2];
h q[4];
cx q[4],q[2];
cx q[0],q[2];
cx q[4],q[2];
h q[2];
h q[4];
cx q[4],q[2];
h q[2];
h q[4];
cx q[4],q[2];
cx q[1],q[2];
h q[2];
cx q[1],q[2];
h q[1];
cx q[4],q[2];
cx q[0],q[1];
h q[2];
h q[4];
h q[1];
cx q[4],q[2];
h q[2];
h q[4];
cx q[4],q[2];
cx q[1],q[2];
cx q[4],q[2];
h q[2];
h q[4];
cx q[4],q[2];
h q[2];
h q[4];
cx q[4],q[2];
sdg q[1];
h q[2];
h q[3];
h q[4];
measure q -> c;
\end{lstlisting}

\clearpage
\section*{References}

\end{document}